\documentclass[aps,prb,twocolumn,showpacs]{revtex4-1}

\usepackage{graphicx}
\usepackage{subfigure}
\usepackage{amsfonts}
\usepackage{amsmath}
\usepackage{amsbsy}

\begin{document}

\title{Magnetotransport and spin dynamics in an electron gas formed at oxide interfaces}
\author{Chenglong Jia and  Jamal Berakdar}

\affiliation{Institut f\"ur Physik, Martin-Luther Universit\"at Halle-Wittenberg,
06099 Halle (Saale), Germany}

\begin{abstract}
We investigate the spin-dependent transport properties of a two-dimensional electron gas formed at  oxides' interface
   in the presence of a magnetic field. We consider several scenarios for the oxides' properties, including oxides with
   co-linear  or spiral magnetic and  ferroelectric order.
   For spiral multiferroic oxides, the magnetoelectric coupling and the topology of the localized magnetic moments introduce additional, electric field controlled spin-orbit coupling that affects the magneto-oscillation of the current.
  An interplay of  this spin-orbit coupling, the exchange field, and of the applied magnetic field results in a quantum, gate-controlled spin and charge Hall conductance.
\end{abstract}
\pacs{75.70.Cn, 73.40.-c, 72.25.-b, 68.47.Gh, 73.20.-r, 85.75.-d}
\maketitle

\section{Introduction}

The transport properties of a semiconductor-based two-dimensional electron gas (2DEG) by external electric and magnetic fields is at the heart of mesoscopic and spintronic research with a wide range of  applications.
A new impetus has been the recent discovery of  2DEG formed at the interface of insulating oxides \cite{Oxides-1,Oxides-2}. This 2DEG can be  laterally confined and patterned to achieve new functionalities such as oxide-based  field effect transistors \cite{Triscone10,JB}. In view of  the
 remarkable phenomena  observed in  the conventional  2DEG under a magnetic field, including  the Shubnikov-de Hass (SdH) effect \cite{SdH,Sander98},  and  the spin/charge Hall effect \cite{Hall-effect} in the presence  of
spin-orbit interactions (SOI) \cite{Rashba,Dresselhaus}, it is timely to consider the properties of oxide-based 2DEG in a magnetic field.
 Generally, the relatively  large effective mass $m^{\ast}$ together with a high carrier concentration $N_e$ at the oxide interfaces (for instance $m^{\ast}/m_{e} \approx 3.2$ with $m_e$ being the free electron mass  and $N_e \approx 5-9 \times10^{13} cm^{-2}$ at the LaAlO$_3$/SrTiO$_{3}$ interface \cite{e-mass}) imply a strong magnetic field for the SdH oscillation and for the quantized Hall conductance to be observable. In addition, oxide interfaces have a multitude of inherent properties such as multiferroicity and strong electronic correlations \cite{RMnO3,ME}. It is our aim here to inspect how such properties affect the magnetotransport in 2DEG.
  One of the exciting results is that, utilizing  2DEG formed at  multiferroic oxides (e.g., RMnO$_{3}$ \cite{TbMnO3,RMnO3}) the magnetotransport  can be  modulated by both a magnetic field and also with a small transverse electric field ($\sim 1kV/cm$). The topology of the local helical magnetic moments in multiferroics results in an induced effective SOI \cite{JB} that is electrically \cite{E-Control} and/or magnetically \cite{H-Control} tunable. Consequently, we find a large magnetic and electric field-dependence of the quantum oscillations in the longitudinal conductance, as well as a spin/charge Hall conductance. Furthermore, the cases of collinear order and/or weak on-site correlations are also considered.

\begin{figure}[b]
\includegraphics[width=0.45\textwidth]{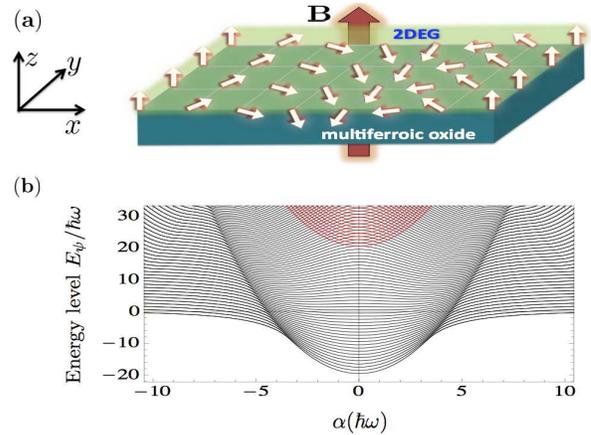}
\caption{(a) Schematic view of 2DEG at the interface of a multiferroic oxide with a polar oxide and in the presence of a magnetic field $\mathbf{B}$. The spiral  $(x-z)$ plane is perpendicular to 2DEG. (b) The  2DEG energy levels as functions of the scaled spin-orbit interaction $\alpha$ with $\Delta /\hbar \omega = 20$.
The spin-down (gray) and the spin-up (black) subbands are separated by the exchange energy.}
\label{fig::system}
\end{figure}

\section{Theoretical formalism}
We start from the general case by considering a 2DEG formed at the intersurface of a spiral multiferroic oxide (e.g., RMnO$_{3}$)  with a polar
oxide (cf. Fig.\ref{fig::system}).
 %
 %
 %
 In a homogeneous, static magnetic field  $\mathbf{B} = (0, 0, B_{z})$ the single-particle  2DEG Hamiltonian reads
\begin{eqnarray}
&& H= h_k + h_{ex}, ~\text{with} \\
&& h_k = \frac{1}{2m^{\ast}} [\mathbf{P} +e \mathbf{A}_{B}]^{2}, ~h_{ex} = U \mathbf{n}_{r}\cdot \boldsymbol{\sigma} + g \mu_{B} \mathbf{B} \cdot \boldsymbol{\sigma}. \nonumber
\label{H1}
\end{eqnarray}
Here  $g$ is the gyromagnetic constant, $\mu_{B}$ is the Bohr magneton, and $\boldsymbol{\sigma}$ are the Pauli matrices.  $U \mathbf{n}_{r}$ is the exchange field
 determined by Coulomb repulsion  and Hund's rule coupling at
 the oxides ionic sites, and can be described by an effective unit-vector field
 $\mathbf{n}_{r}$ and  a strength  $U$ .  A useful parameterizations is $$\mathbf{n}_{r} = (\sin \theta_{r},0,\cos \theta_{r})$$ with $$\theta_{r} = \mathbf{q}_{m} \cdot \mathbf{r}$$ and $\mathbf{q}_{m} = (q,0,0)$ is the spin wave vector of the spiral \cite{TbMnO3,E-Control}.
 $$\mathbf{A}_{B} = ( -B_{z} y, 0, 0)$$ in the Landau gauge.  The case of a collinear order and/or weak electronic
 correlations follow as special cases, as discussed below.  Upon a unitary gauge transformation $$U_{g} = exp(-i\theta_{r} \sigma_{y}/2)$$  the local quantization axis becomes  aligned with $\mathbf{n}_{r}$  on the expense of   introducing the topological vector potential
 $$ \mathbf{A}_{g} = -i\hbar U^{\dag}_{g} \boldsymbol{\nabla} U_{g} = (- \hbar q \sigma_{y}/2, 0, 0).$$
 The transformed kinetic energy \cite{JB} reads $$\tilde{h}_k = \frac{1}{2m^{\ast}} [\tilde{\mathbf{P}} +e \mathbf{A}_{B} + \mathbf{A}_g]^{2}$$ (hereafter, transformed quantities are marked by a tilde).  $\mathbf{A}_{g}$ depends only on the geometry of the local magnetization at the oxide and acts as a  helicity $q$ and momentum-dependent effective SOI  $\sim q\tilde{P}_{x}\tilde{\sigma}_{y}$.
 The transformed term $\tilde{h}_{ex}$ turns however diagonal, $\tilde{h}_{ex} = \Delta \tilde{\sigma}_z$ (after omitting a small periodic modulation on the spins \cite{footnote}).
 The value of   $\Delta$ depends on the system under study: \\
 \emph{i)} For a strong local correlation compared to the
 external Zeeman field we find $\Delta \approx U$. This is the case of spiral
  multiferriocs.\\
   \emph{ii) }The case of a collinear $\mathbf n(\mathbf r)$ (no ferroelectric order) corresponds to
   $q\to 0$, and hence $\mathbf{A}_g \rightarrow 0$.\\
  \emph{iii)} For a weak local correlation field $U\to 0$ and/or under a strong magnetic field
  we find  $\Delta \approx g \mu_{B} B_z$, i.e.  the conventional 2DEG in
   a magnetic field. \\
   The cases \emph{i)}- \emph{iii) } correspond to different combinations of the oxides at which interface
    the 2DEG is formed. We will focus here on  \emph{i)} and \emph{ii)}. Predictions for the case \emph{iii)} are readily deduced, the physics
     however in this case resembles the well-known situation in a semi-conductor-based 2DEG (without SOI).\\
   For the transformed system we infer $$[\tilde{P}_{x}, \tilde{h}_k + \tilde{h}_{ex} ] =0.$$ Introducing  the bosonic operator
\begin{eqnarray}
a = [\tilde{P}_{x} - eB_{z}y  - i \tilde{P}_{y}]/ \sqrt{2m\hbar \omega}
\end{eqnarray}
and scaling all energies in unit of the cyclotron energy $$\hbar \omega = \hbar eB_{z}/m^{\ast},$$ we obtain the dimensionless harmonic Hamiltonian
\begin{equation}
\bar{H} = (a^{\dag}a +1/2) -\alpha \tilde{\sigma}_{y} (a^{\dag} + a)  +\alpha^{2} + \bar{\Delta} \tilde{\sigma}_{z}
\label{H3}
\end{equation}
where
$$\bar \Delta = \Delta /\hbar \omega,\;\; \alpha = \frac{q l_{B}}{2\sqrt{2}\bar a}, \;\; l_{B} = \sqrt{\hbar/eB_{z}}. $$ $\bar a$ is the lattice constant, and $l_{B}$ is the magnetic length.

In the limit of a collinear spin order, the eigenstates of the Hamiltonian are the Landau levels, $$|n,s \rangle = \frac{(a^{\dag})^{n}}{\sqrt{n!}}|0\rangle \otimes |s \rangle$$ with the energy $$\epsilon_{ns} = (n +1/2) + s \bar \Delta +\alpha^{2},\; n=0,1,2,\cdots,\quad  s = \pm.$$ However,  in most cases,  the state $|n,s \rangle$ is coupled to $|n\pm 1 ,\bar s\rangle$ via  SOI. The case
  of a weak exchange (local correlation) field follows for $\bar{\Delta}=0$ (cf.~[\onlinecite{footnote}]).
  Expanding the eigenstate $$|\psi \rangle= \sum_{n\sigma}C_{n}^{s}|n,s \rangle$$ in the Landau space, the secular equation $$\bar{H} |\psi \rangle = E_{\psi} |\psi \rangle$$ leads to the following matrix form,
\begin{eqnarray}
&& E_{\psi} \left[ \begin{array}{c} ... ~
C_{n-1}^{\bar s} ~ C_{n}^{s} ~ C_{n+1}^{\bar s} ~... \end{array} \right]^{T} =  \\
&& \left[\begin{array}{ccccc}... & ... & ... & ... & ... \\
... &  \epsilon_{n-1,\bar s} & i \alpha \bar{s} \sqrt{n} & 0 & ...
\\... & i \alpha s \sqrt{n} & \epsilon_{n,s}  &  i \alpha s \sqrt{n+1} & ...
\\... & 0 & i \alpha \bar{s} \sqrt{n+1}  & \epsilon_{n+1,\bar s}   & ...
\\... & ... & ... & ... & ...\end{array}\right] \left[ \begin{array}{ccccc} ... \\
C_{n-1}^{\bar s} \\ C_{n}^{s} \\ C_{n+1}^{\bar s}  \\... \end{array} \right]  . \nonumber
\end{eqnarray}
The Hamiltonian in the Landau space reduces then to two independent, infinitely dimensional tridiagonal matrices with reference to two groups\cite{LLs} $| n,(-1)^n \rangle$ and $| n, -(-1)^n \rangle$. The energy spectrum can be obtained numerically by truncating the matrix dimensions while including a sufficient number of Landau levels (up to 1000 energy levels were taken during the calculations of the transport coefficients). \\
 Fig.\ref{fig::system}b  shows the energy levels as a function of the SOI $\alpha$ with $\bar{\Delta} = 20$.  Two spin subbands (parallel or antiparallel to the exchange field) are shown.   Hence, the electronic structure   varies strongly with $\alpha$ which is tunable by a small transverse electric field due to the magnetoelectric coupling \cite{E-Control}. The high density of states around $\alpha \approx 5$ follows from  a perturbation considerations:  \\
 Let SOI be  smaller than the exchange splitting, $\alpha \ll \bar \Delta$. Up to a second order in $\alpha$ the two energy branches read
\begin{eqnarray}
E_{ns} (\alpha) = \epsilon_{ns}+ s \frac{(n+1) \alpha^2}{ 2\bar \Delta - s} + s \frac{n \alpha^2 }{2\bar \Delta +s}.
\end{eqnarray}
At critical value $$\alpha_{c} = [(4\bar{\Delta}^{2}-1)/4 \bar{\Delta}]^{1/2},$$ we find  infinitely degenerate states, $E_{ns=-1}(\alpha_{c}) \equiv 0$.  For large $\alpha$ and/or $n$, the degeneracy is lifted and distributed around  $\alpha_{c}$, e.g.  $\alpha_c = 4.47 $ with a given $\bar{\Delta} =20$.
\begin{figure}[t]
\includegraphics[width=0.45\textwidth]{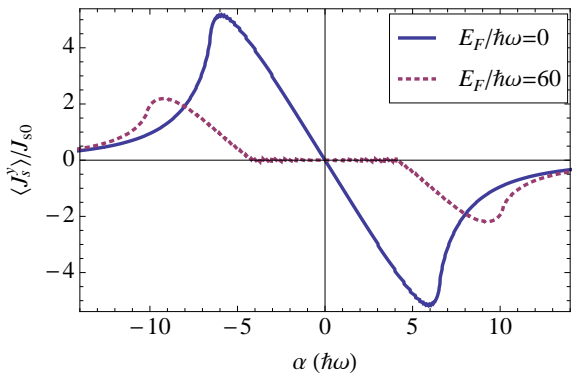}
\caption{The persistent spin current $\langle \hat{J}_{s}^{y} \rangle$  in units of  $J_{s0}=\sqrt{\hbar \omega / 2m}$ vs.~SOI strength $\alpha$  with $\Delta /\hbar \omega =43$. Solid line:  $E_{F} < \Delta$,  only the spin-down subband intersects  the Fermi level. Dotted line:  $E_{F} > \Delta$,  both  spin subbands are occupied when $|\alpha | \lesssim 4$. }
\label{fig::Js}
\end{figure}

\section{Persistent spin current}
As pointed out in Ref.[\onlinecite{JB}] without a magnetic field, a nonzero spin current $\langle J_{s}^{y} \rangle$ is  generated when only the spin-down subband is below the Fermi level $E_F$.  With a magnetic field $B_{z}$ the spin current is
\begin{equation} \langle \hat{J}_{s}^{y} \rangle =1/ \nu \sum_{\psi} \langle \psi |\hat{J}_{s}^{y} |\psi \rangle f(E_{\psi}), \end{equation}
where $\nu$ is the filling factor and $f$ is the Fermi distribution function.  The spin current operator $\hat{J}_s^y$  for the $y$ component of the spin is defined as
\begin{eqnarray}
&& \hat{J}_{s}^{y} = {1 \over 2} (\tilde{\sigma}_{y} \hat{v}_{x} + \hat{v}_{x} \tilde{\sigma}_{y}),  \\
&& \hat{v}_x = \frac{[x, \bar{H}]}{i\hbar} = \sqrt{\frac{\hbar \omega}{2m}} [(a^{\dag} + a) - 2 \alpha \tilde{\sigma}_y].
\end{eqnarray}
Note, $\tilde{\sigma}_y = \sigma_y$ because $[\sigma_y, U_g] \equiv 0$. An overall $\alpha$-dependence of the amplitude of spin current is related to the electron density through the filling factor (Fig.\ref{fig::Js}).  In the case of $E_{F} / \hbar \omega = 60$, two subbands are both occupied only when $|\alpha | \lesssim 4$. It is then numerically confirmed that the persistent spin current vanishes when two spin subbands are  intersected by the Fermi level \cite{JB}.  No charge current is generated, i.e. $\langle \hat {J}_{cx} \rangle = \langle -e \hat{v}_x \rangle=0$ in the absence of an in-plane electric field. The spin polarization vanishes as well $\langle \sigma_{y} \rangle =0$ under a normal magnetic field; only a pure, electrically tunable persistent spin current exists.
\setcounter{subfigure}{0}
\begin{figure}[t]
\includegraphics[width=.5\textwidth]{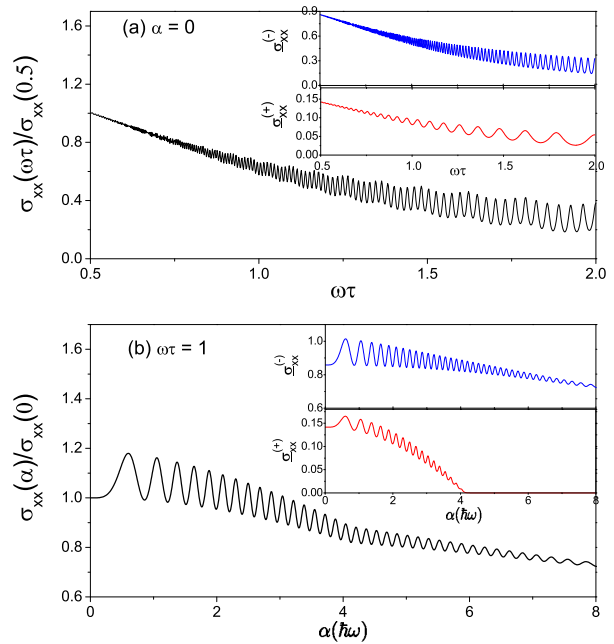}\caption{The longitudinal conductivity $\sigma_{xx}$ as functions of (a) magnetic field $\omega \tau$ and (b) SOI strength  $\alpha$ at a fixed Fermi energy $E_F \tau /\hbar = 60$. The exchange energy is $\Delta \tau /\hbar =43$. Insets show the contributions from the different subband, $\underline{\sigma}_{xx}^{(\pm)}$ is re-scaled with respected to the total conductivity $\sigma_{xx}$ at (a) $\omega \tau =0.5$ and (b) $\alpha=0$.}
\label{fig::sxx}
\end{figure}

\section{Longitudinal conductivity}
The topological SOI is exactly analogous to the semiconductor-based 2DEG with the Rashba and Dresselhaus spin-orbit couplings being equal in strength.
Therefore, the spin beats in the magneto-oscillation at the helimagnetic interface are completely suppressed \cite{SdH-2}.  However, the oscillatory magnetoresistance contains at least two components:  SdH oscillations of the spin-down and spin-up subbands, respectively \cite{MIS1}.   $\bar{\Delta} \gg 1$ means a large energy separation between the two subbands, $\sim 2\langle \tilde{\sigma}_z \rangle \bar{\Delta}$, which results in  negligible off-diagonal terms of the  Green function \cite{SCBA}.  The longitudinal conductivity is then given by $ \sigma_{xx} = \sigma_{xx}^{(-)} + \sigma_{xx}^{(+)}$ with
\begin{eqnarray}
\sigma_{xx}^{(\pm)} = \frac{m\omega}{4\pi^2} \sum \frac{ \langle \psi_{\pm} | \hat{J}_{cx} | \psi^{\prime}_{\pm} \rangle \langle \psi^{\prime}_{\pm} | \hat{J}_{cx} | \psi_{\pm} \rangle }{(E_{\psi_{\pm}} - \mu_{\pm} + i\hbar/\tau_{\pm})(E_{\psi'_{\pm}} - \mu_{\pm} + i\hbar/\tau_{\pm})}\nonumber
\end{eqnarray}
where $\langle \psi_{\pm} | \hat{J}_{cx} | \psi^{\prime}_{\pm} \rangle$ is the matrix elements of the charge current density operator.  $\mu_{\pm}$ is the energy distance between the Fermi energy and the bottom of the spin-up(down) subband. $\tau_{\pm}$ is the total relaxation time including the intrasubband and intersubband scattering at a short-range random potential.
%
 Experiments on semiconductor-based   2DEG \cite{LorenzianWidth}
 suggest a  Lorentzian shape as an excellent fit for the disorder-induced broadening of Laudau levels; we assume this to hold in our case, i.e.
 $\tau_{\pm} = \tau$.
  Fig.\ref{fig::sxx}a demonstrates the dependence of $\sigma_{xx}$ on the magnetic field for the following parameters: $E_F \tau /\hbar = 60$, $\Delta \tau/\hbar = 43$, and $\alpha =0$. The spectrum of $\sigma_{xx}$ consists of two harmonics: The oscillating part of the \emph{partial} conductivities $\sigma_{xx}^{(\pm)}$ is  described well analytically by $\cos ( 2\pi \frac{\mu_{\pm}}{\hbar \omega} +\pi )$.  The magnitudes are proportional to the carrier concentrations in the spin-up and spin-down subbands, respectively.  On the other hand, assuming a fixed Fermi energy the carrier concentration varies with the electrically tunable spin-orbit interaction parameter $\alpha$,  the quantum oscillation of  the longitudinal conductance $\sigma_{xx}$ is caused then
  by a  crossing of the chemical potential and the energy levels, see also in Fig.\ref{fig::sxx}b.
  The harmonics stem from a change of the energy distance $\mu_{\pm}$, which induces an equivalent frequency of $\sigma_{xx}^{(\pm)}$. As $\alpha > 4$,   only the low subband is occupied however, $\sigma_{xx}^{(+)} \rightarrow 0$.

\setcounter{subfigure}{0}
\begin{figure}[b]
\includegraphics[width=.5\textwidth]{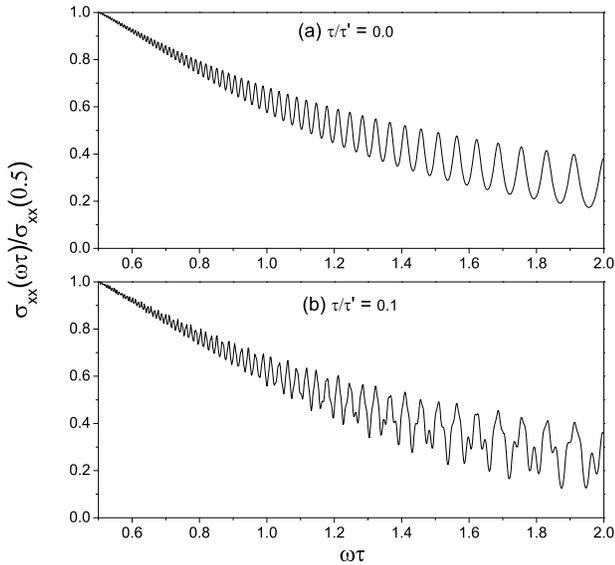}\caption{The longitudinal conductivity $\sigma_{xx}$ vs. magnetic field $\omega \tau$ for various intersubband  scattering intensities: (a) $\tau/\tau^{\prime} =0.0$, (b) $\tau/\tau^{\prime} =0.1$ at a fixed Fermi energy $E_F \tau /\hbar = 0$. The exchange energy is $\Delta \tau /\hbar =43$ and the SOI $\alpha =0$.}
\label{fig::MIS}
\end{figure}

 When  accounting explicitly for intersubband transitions,  the total relaxation time $\tau_{\mp}$ reads \cite{MIS2}
\begin{eqnarray}
\tau_{-} = \tau [ 1+ ( 1- \frac{\tau}{\tau^{\prime}} ) F_{-} + \frac{\tau}{\tau^{\prime}} F_{+} ]  \\
\tau_{+} = \tau [ 1+ ( 1- \frac{\tau}{\tau^{\prime}} ) F_{+} + \frac{\tau}{\tau^{\prime}} F_{-} ]
\end{eqnarray}
with the times $\tau$ and $\tau^{\prime}$ for intra- and inter-subband scattering at a short-range random potential, respectively. $F_{\pm}$ is oscillatory as inferred from

\begin{equation}
F_{\pm} = 2 \cos ( 2\pi \frac{\mu_{\pm}}{\hbar \omega} +\pi ) \exp(-\frac{\pi}{\omega \tau})
\end{equation}
to the first order of the parameter $\exp(-\frac{\pi}{\omega \tau})$. Fig.\ref{fig::MIS} presents the dependencies of the $\sigma_{xx}$ on the magnetic field under the condition that only the lower subband is below the Fermi level: $E_F \tau /\hbar =0$ and $\Delta \tau/\hbar =43$. A high-frequency oscillation component emerges (termed  magneto-intersubband scattering (MIS) effect \cite{MIS1}),  which behaves as $\cos (4 \pi \frac{\langle \tilde{\sigma}_z \rangle \Delta}{\hbar \omega})$ and is distinguishable from  the background harmonic $\cos ( 2\pi \frac{\mu_{-}}{\hbar \omega} +\pi )$. Upon increasing the temperature,  MIS   dominates the oscillatory magnetoresistance due to its weak-temperature damping compared to SdH oscillations \cite{Sander98}.

\setcounter{subfigure}{0}
\begin{figure}[t]
\includegraphics[width=0.45\textwidth]{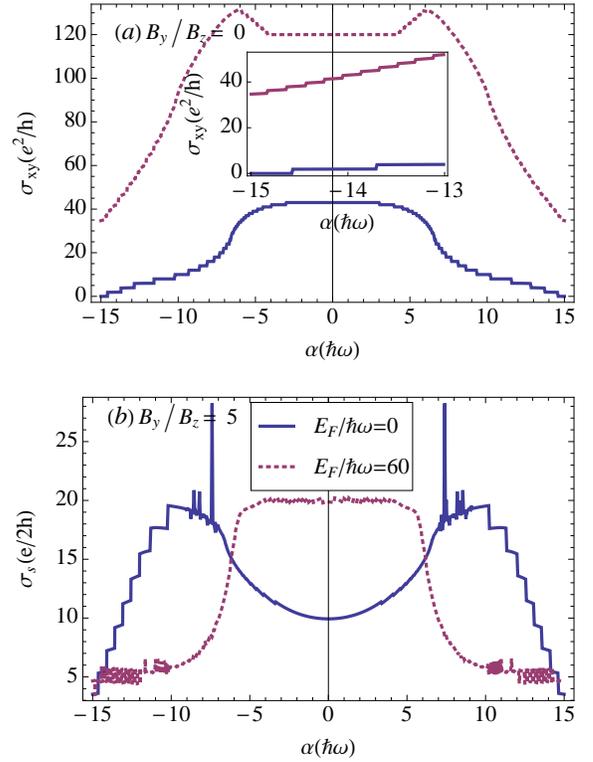}
\caption{(Top): The charge Hall conductivity, $\sigma_{xy}$ vs.  SOI parameter $\alpha$ with no in-plane component of the magnetic field ($B_{y}/B_{z} =0$).  Inset:  a close-up of the quantization of $\sigma_{xy}$.  (Bottom):  The spin Hall conductivity, $\sigma_{s}$ under $B_{y}/B_{z} = 5$. The exchange energy is $\Delta /\hbar \omega =43$. }
\label{fig::Hall}
\end{figure}

\section{Hall effect}
Following Rashba \cite{sum-rules}, the charge(spin) Hall conductivity is  given by  Kubo-Greenwood formula,

\begin{equation}
\sigma_{xy,s} = -\frac{e}{\pi l_{B}^{2} }\sum' \frac{\Re \langle \psi | \hat{y} |\psi^{\prime} \rangle \langle \psi^{\prime} |\hat{J}_{cx,s} |\psi \rangle }{E_{\psi^{\prime}} - E_{\psi} }.
\label{Kubo}
\end{equation}
The prime over the sum indicates a summation only over the state $|\psi^{\prime} \rangle$ below the Fermi level ($ E_{\psi^{\prime}} < E_{F}$)  and  $|\psi \rangle$ above the Fermi level ($E_{\psi} > E_{F}$). The numerical evaluation shows a quantization of the charge Hall conductance,  $\sigma_{xy} = v e^{2}/ 2 \pi \hbar$ (see in Fig.\ref{fig::Hall}a),  for the charge current carried by each state is not changed by SOI \cite{Hall-effect}.  A similar  electric-field effect on the charge Hall conductivity has been reported in Graphene p-n junctions \cite{Graphene}, in which the carrier type and density can be controlled by external gates as well. Different from a semiconductor 2DEG, where a resonant spin Hall conductance is predicated for equal Rashba and Dresselhaus SOI\cite{R-SHC}, the spin Hall conductivity is zero in our system in the presence of only a perpendicular magnetic field due to the absence of a crossing of any adjacent energy-levels.

When the magnetic field is tilted in the $y-z$ plane with respected to the $z$ axis, the in-plane $y$ component $B_{y}$ induces a nonzero spin polarization $\langle \sigma_{y} \rangle =1/\nu \sum_\psi \langle \psi | \hat{\sigma}_y |\psi \rangle$, which renders the oxide magnetic order not exactly coplanar.  A nonzero Berry's curvature is expected in the momentum space \cite{Berry-Phase}, and so is the spin Hall conductivity $\sigma_s$.  The $\alpha$-dependence of $\sigma_s$ is different from that of $\sigma_{xy}$ (see Fig.\ref{fig::Hall}b), which implies that the spin Hall conductance is not totally dominated by the spin-polarized charge current.  A step-like structure is found in the spin-Hall conductivity for $E_{F} <  \Delta $. It should be noted that the filling factor $\nu$, the persistent spin current $\langle J_s^y \rangle$, and the charge Hall conductivity $\sigma_{xy}$ are not sensitive to the in-plane magnetic field $B_y$. However, the competition between the Zeeman energy and the SOI results in  sharp changes of the spin polarization  $\langle \sigma_{y} \rangle$, which gives rise to resonant peaks around the degenerate point $\alpha_c$ (for $E_F =0$) and to a quick oscillations (for $E_F =60 \hbar \omega$) of the spin Hall conductivity $\sigma_{s}$ .

\section{Summary}
In summary, we  studied the spin current, the magneto-oscillations, and the spin/charge Hall effect in 2DEG at oxide interfaces exposed to  magnetic fields.  The transport behavior is determined by an interplay between the exchange interaction, the magnetic field, and the effective SOI induced by the spiral geometry of the local magnetic order. The electrically tunable topological SOI allows a control of the  system's magnetotransport by a small transverse electric field. Our predictions are accessible experimentally. In fact,
very recently,  the SdH effect has been experimentally observed for SrTiO$_3$/LaAlO$_3$ \cite{SdH-in-oxides}, where an intense magnetic field up to 31.5T was applied.  A collinear spin phase is thus achieved, and the spin helicity $q \rightarrow 0$, which corresponds to a special case described by the Hamiltonian Eq.\ref{H3} with $\alpha =0$ and $\Delta = g\mu_B B_z$.


\acknowledgments
We thank Zhen-Gang Zhu,  Ionela Vrejoiu, and Georg Schmidt   for valuable discussions.
 This research is supported by the DFG (Germany) through the project-B7-  in the SFB762: {\it functionality of oxide interfaces}.


\begin{thebibliography}{99}
\bibitem{footnote} The transformed exchange energy is $\tilde{h}_{ex} =U \tilde{\sigma}_{z} + g\mu_{B}B_{z}\mathbf{n}_{r}^{g} \cdot \boldsymbol{\tilde{\sigma}}$ with $\mathbf{n}_{r}^{g} = (-\sin \theta_{r}, 0, \cos \theta_{r})$. For multiferroics with spiral spins, the applying magnetic field $\mathbf{B}$ should be  as low as to preserve the local magnetic order, e.g.  $B<4T$ for  experiments on Eu$_{0.55}$Y$_{0.45}$MnO$_{3}$ \cite{H-Control}.  The Zeeman splitting in $\tilde{h}_{ex}$ is  subsidiary  as compared to  the exchange field. For  strong magnetic fields however, e.g. up to 31.5T in the experiment on SrTiO$_3$/LaAlO$_3$  \cite{SdH-in-oxides} a collinear spin phase is achieved. The relatively {weak} exchange interaction  provides then a one-dimensional periodic modulation on the spins.
\end{thebibliography}
\end{document}